\RequirePackage{fix-cm}
\documentclass[smallextended]{svjour3}
\smartqed
\usepackage[T1]{fontenc}
\usepackage{times,graphicx,amsmath,amssymb,amsfonts,mathrsfs,mathtools,enumerate,cite,float,marvosym,geometry,needspace,placeins}
\usepackage[colorlinks,bookmarksopen,bookmarksnumbered,citecolor=blue,linkcolor=blue,urlcolor=blue]{hyperref}
\DeclareFontFamily{U}{rsfs}{\skewchar\font127}
\DeclareFontShape{U}{rsfs}{m}{n}{<-6.5> rsfs5 <6.5-7.5> rsfs7 <7.5-> rsfs10}{}
\mathtoolsset{showonlyrefs}
\hypersetup{pdftitle={Construction and regularity of lump solutions of the modified KP-I equation},pdfauthor={Tianwei Qiu and Zhen Wang}}
\graphicspath{{figures/}}
\allowdisplaybreaks[2]
\newcommand{\ii}{\mathrm i}
\newcommand{\R}{\mathbb R}
\newcommand{\C}{\mathbb C}
\newcommand{\dd}{\,\mathrm d}
\newcommand{\im}{\operatorname{Im}}
\newcommand{\re}{\operatorname{Re}}
\newcommand{\diag}{\operatorname{diag}}
\newcommand{\sgn}{\operatorname{sgn}}
\newcommand{\panel}[3]{\begin{minipage}{#1\linewidth}\centering\includegraphics[width=\linewidth]{#2}\par\small #3\end{minipage}}

\begin{document}
\begin{sloppypar}
\title{Construction and regularity of lump solutions of the modified KP-I equation}
\titlerunning{Construction and regularity of lump solutions of the mKP-I equation}
\author{Tianwei Qiu \and Zhen Wang}
\authorrunning{T. Qiu and Z. Wang}
\institute{T. Qiu \at School of Mathematical Sciences, Beihang University, Beijing 100191, China\\
\and Z. Wang \Letter \at School of Mathematical Sciences, Beihang University, Beijing 100191, China\\\email{wangzmath@163.com}}
\date{}
\setlength{\headerboxheight}{0pt}
\renewcommand{\makeheadbox}{}
\maketitle

\begin{abstract}
We express the $\tau$-functions of standard and higher-order lump solutions of the modified Kadomtsev--Petviashvili-I (mKP-I) equation as Gram determinants in integral form. Unlike the KP-I equation, the mKP-I $\tau$-function is complex-valued, and its nonvanishing does not follow from the positive definiteness of a Hermitian Gram matrix. We derive explicit spectral conditions for the regularity of these solutions; for higher-order lumps, the sharp condition depends only on the spectral parameter and the largest entry of the index vector. We also prove that the lump solutions and all their spatial derivatives decay algebraically in every spatial direction, uniformly on compact time intervals.
\end{abstract}

\section{Introduction}\label{sec:intro}

The Kadomtsev--Petviashvili (KP) equation is a classical model for weakly nonlinear dispersive waves with slow transverse variation \cite{KadPet1970}. The KP and modified KP (mKP) equations are related by a two-dimensional Miura transformation \cite{Konopelchenko1982,DubKonMiura}. In this paper, we consider the mKP-I equation
\begin{equation}\label{eq:mkp}
 v_t+v_{xxx}+\tfrac32v^2v_x-3\partial_x^{-1}v_{yy}+3v_x\partial_x^{-1}v_y=0,
\end{equation}
where $(\partial_x^{-1}g)(x,y,t)=\int_{-\infty}^x g(x',y,t)\dd x'$.

Lumps are rational solutions localized in all spatial directions. Explicit lump solutions of the KP-I equation were first obtained by Manakov et al.~\cite{ManakovEtAl1977}, and standard lumps were constructed by Ablowitz and Satsuma via the long-wave limit of soliton solutions \cite{AblowitzSatsuma1978,SatsumaAblowitz}. Higher-order (multi-pole) lumps were subsequently derived via the inverse scattering transform (IST) \cite{FokasAblowitz1983,VillarroelAblowitz1999} and binary Darboux transformations \cite{AblowitzEtAl2000}. Standard lumps with distinct velocities undergo normal scattering, whereas higher-order lumps share a common asymptotic velocity and undergo anomalous scattering \cite{SatsumaAblowitz,Gorshkov1993}. In Hirota's bilinear framework, Pelinovsky constructed higher-order rational solutions of the KP hierarchy \cite{Pelinovsky1994}, Chakravarty and Zowada classified higher-order KP-I lump patterns via Gram determinants and Schur polynomials \cite{ChakravartyZowada2022,ChakravartyZowada2023}, and Yang and Yang characterized their long-time patterns through the roots of the Yablonskii--Vorob'ev and Wronskian--Hermite polynomials \cite{YangYang2022}. For the KP-I equation, the lump $\tau$-function is the determinant of a positive-definite Hermitian Gram matrix, which immediately guarantees that it has no real zeros and that the solution is regular for all spectral parameters off the real axis \cite{ChakravartyZowada2022}.

For the mKP equation, Gesztesy and Schweiger derived Wronskian rational solutions of the mKP-II equation \cite{GesztesySchweiger}, Konopelchenko and Dubrovsky constructed standard mKP-I lumps via the IST \cite{KonopDubrovsky}, and Yong et al.~\cite{YongEtAl2020} obtained the fundamental mKP-I lump by Hirota's bilinear method. In \cite{QWY}, we derived standard and second-order lump solutions of \eqref{eq:mkp} via long-wave limits, and we further applied the generalized long-wave limit method \cite{GLWLKP} to obtain second- and third-order lump solutions in \cite{GLWLmKP}.

The KP and mKP hierarchies admit bilinear forms in terms of adjacent $\tau$-functions \cite[Appendix~1]{JimboMiwa1983} and are connected through generalized Hirota bilinear identities \cite{BogdanovKonopelchenko1998}. Unlike the KP-I case, the bilinear form of the mKP-I equation involves a complex-valued $\tau$-function $f$, whose nonvanishing in $\R^3$ does not follow from the positive definiteness of a Hermitian matrix; indeed, $f$ can vanish in $\R^3$ for certain spectral parameters \cite{QWY}. Moreover, the nonlocal operator $\partial_x^{-1}$ in the nonlinear term $3v_x\partial_x^{-1}v_y$ requires sufficient spatial decay as $x\to-\infty$ for $v$ to satisfy \eqref{eq:mkp}.

In this paper, we construct general higher-order lump solutions of \eqref{eq:mkp} associated with arbitrary index vectors via the bilinear mKP hierarchy and express the $\tau$-functions of both standard and higher-order lumps as Gram determinants in integral form. We derive explicit spectral conditions ensuring that the $\tau$-functions have no zeros in $\R^3$. We also prove that the regular lump solutions and all their spatial derivatives decay algebraically in every spatial direction, ensuring that the nonlocal terms in \eqref{eq:mkp} are well defined.

\subsection{Main results}\label{sec:results}

In terms of a complex-valued $\tau$-function $f$, Eq.~\eqref{eq:mkp} admits the bilinear form \cite{QWY}
\begin{equation}\label{eq:bilinear}
 \begin{aligned}
 &v=2\ii\left(\log\frac{f^*}{f}\right)_x=4\im\frac{f_x}{f},\\
 &(D_y+\ii D_x^2)f^*\cdot f=0,\qquad (D_t+D_x^3+3\ii D_xD_y)f^*\cdot f=0.
 \end{aligned}
\end{equation}
Throughout, $*$ denotes complex conjugation, ${}^{\mathsf T}$ the transpose, and ${}^\dagger$ the conjugate transpose (or adjoint). Hirota's bilinear operators in \eqref{eq:bilinear} are defined by \cite{Hirota2004}
\[
 D_x^mD_y^nD_t^k g\cdot f=\left.(\partial_x-\partial_{x'})^m(\partial_y-\partial_{y'})^n(\partial_t-\partial_{t'})^k g(x,y,t)f(x',y',t')\right|_{(x',y',t')=(x,y,t)}.
\]
The $\tau$-functions $f$ constructed below are polynomials in $(x,y,t)$. Whenever $f$ has no zeros in $\R^3$, the function $v$ defined by \eqref{eq:bilinear} is smooth on $\R^3$; if $v$ also satisfies Eq.~\eqref{eq:mkp}, we call $v$ a regular solution.

Both classes of solutions arise from Gram determinants via differentiation with respect to spectral parameters and a complex conjugation reduction (see Sect.~\ref{sec:construction}). Standard lumps use distinct conjugate pairs of spectral parameters, whereas higher-order lumps arise from Taylor expansions at a single conjugate pair. For $\boldsymbol s=(s_1,s_2,\ldots)$, the elementary Schur polynomials $S_j(\boldsymbol s)$ are defined by the generating function \cite{Macdonald1995}
\begin{equation}\label{eq:schurseries}
 \exp\left(\sum_{k\ge1}s_k\epsilon^k\right)=\sum_{j\ge0}S_j(\boldsymbol s)\epsilon^j,\qquad S_j=0\quad(j<0).
\end{equation}

\begin{theorem}[Lump solutions of the mKP-I equation]\label{th:construction}
The following two classes of polynomial $\tau$-functions satisfy the bilinear equations in \eqref{eq:bilinear}:

\smallskip
\noindent\textup{(i)} Standard lump solutions \cite{QWY}. For a positive integer $M$, let $\lambda_j=a_j+\ii b_j$ ($a_j\in\R$, $b_j>0$, $1\le j\le M$) be distinct spectral parameters and $\gamma_j\in\C$ be phase parameters. Setting $\lambda_{M+j}=\lambda_j^*$ and $\gamma_{M+j}=\gamma_j^*$, one has
\begin{equation}\label{eq:standard}
 \begin{aligned}
  f_{{\rm std},M}&=\det\mathsf A,\qquad 1\le j,k\le 2M,\\[3pt]
  \mathsf A_{jk}&=\begin{cases}
   x+\lambda_jy+3\lambda_j^2t+\dfrac{\ii}{\lambda_j}+\gamma_j, & j=k,\\[8pt]
   \dfrac{2\ii}{\lambda_j-\lambda_k}, & j\ne k.
  \end{cases}
 \end{aligned}
\end{equation}

\smallskip
\noindent\textup{(ii)} Higher-order lump solutions. Let $\lambda=a+\ii b$ ($a\in\R$, $b>0$) be a spectral parameter, and set $\mu=\ii\lambda^*/2$ and $q=\mu^*$. Let $N$ be a positive integer, $\Lambda=(n_1,\ldots,n_N)$ be a strictly increasing index vector of positive integers with largest entry $d=n_N$, and $\zeta_1,\ldots,\zeta_d\in\C$ be phase parameters (with $\zeta_k=0$ for $k>d$). Define
\begin{equation}\label{eq:physicalschur}
 \begin{aligned}
  s_1&=x-2\ii\mu y-12\mu^2t+\zeta_1,\qquad
  &s_2&=-\ii y-12\mu t+\zeta_2,\\[2pt]
  s_3&=-4t+\zeta_3,\qquad
  &s_k&=\zeta_k\quad(k\ge4).
 \end{aligned}
\end{equation}
Then
\begin{equation}\label{eq:highertau}
 \begin{aligned}
  f_\Lambda&=\det(\mathsf P\mathsf H\mathsf P^\dagger),\qquad
  \mathsf P_{ik}=S_{n_i-k}(\boldsymbol s),\\[3pt]
  \mathsf H_{k\ell}&=[\epsilon^k\delta^\ell]\frac{q+\delta}{b+\epsilon+\delta}
  \qquad(1\le i\le N,\ 0\le k,\ell\le d),
 \end{aligned}
\end{equation}
where $[\epsilon^k\delta^\ell]$ extracts the coefficient of $\epsilon^k\delta^\ell$ in the bivariate Taylor expansion of $(q+\delta)/(b+\epsilon+\delta)$ at $(\epsilon,\delta)=(0,0)$.
\end{theorem}

Our next theorem gives explicit spectral conditions ensuring that the $\tau$-functions in Theorem~\ref{th:construction} have no zeros in $\R^3$ and yield regular, algebraically decaying solutions of \eqref{eq:mkp}. Throughout, we set $R=(x^2+y^2)^{1/2}$.

\begin{theorem}[Regularity of lump solutions]\label{th:regularity}
Suppose that either of the following conditions holds for the lump solutions in Theorem~\ref{th:construction}:

\smallskip
\noindent\textup{(i)} For the standard lump solution in \eqref{eq:standard}, $a_1,\ldots,a_M$ are all positive or all negative, and
\begin{equation}\label{eq:standardarctan}
 \sum_{j=1}^M\arctan\frac{b_j}{|a_j|}<\frac{\pi}{2}.
\end{equation}

\smallskip
\noindent\textup{(ii)} For the higher-order lump solution in \eqref{eq:highertau},
\begin{equation}\label{eq:cotbound}
 |a|>b\cot\frac{\pi}{d+1}.
\end{equation}

\smallskip
\noindent Then for any fixed phase parameters, the corresponding $\tau$-function has no zeros in $\R^3$, and $v=4\im(f_x/f)$ is a regular solution of \eqref{eq:mkp}; as $R\to\infty$, $\partial_x^j\partial_y^k v=O(R^{-2-j-k})$ for all integers $j,k\ge0$, uniformly on compact time intervals.
\end{theorem}

To prove Theorem~\ref{th:regularity}, we first establish quadratic-form and spectral estimates for $\partial_s$ in Sect.~\ref{sec:tools}, and then prove in Sect.~\ref{sec:decay} that any nonvanishing $\tau$-function whose leading homogeneous part in $(x,y)$ has constant argument yields an algebraically decaying solution of Eq.~\eqref{eq:mkp}. In Sect.~\ref{sec:regproofs}, we combine these results with the Gram determinant representations from Sect.~\ref{sec:construction} to show that the $\tau$-functions are bounded away from zero under \eqref{eq:standardarctan} and \eqref{eq:cotbound} and satisfy the leading-term conditions.

The remainder of the paper is organized as follows. Section~\ref{sec:construction} derives the determinant representations for both classes of lump solutions. Section~\ref{sec:regularity} proves Theorem~\ref{th:regularity}. Section~\ref{sec:visualization} illustrates the fundamental lump and the evolution of higher-order lump solutions.

\section{Determinant representations of general lump solutions}\label{sec:construction}

\subsection{Bilinear identities for Gram determinants}\label{sec:adjacent}

Both classes of lump solutions are constructed from Gram determinants in the mKP hierarchy (see \cite{JimboMiwa1983,Hirota2004}). We first record the differential and shift relations needed to derive the bilinear equations.

\begin{lemma}\label{lem:adjacent}
Let $\mathsf M^{(n)}=(m_{ij}^{(n)})$ be an $N\times N$ matrix and $\boldsymbol\phi^{(n)},\boldsymbol\psi^{(n)}$ be $N$-component column vectors whose entries are smooth functions of $x_1,x_2,x_3$. Denote partial derivatives with respect to $x_1,x_2,x_3$ by subscripts $1,2,3$, and suppose that for all $n\in\mathbb Z$ these quantities satisfy the differential and shift relations
\begin{equation}\label{eq:gramflows}
 \begin{aligned}
  \mathsf M^{(n)}_1&=\boldsymbol\phi^{(n)}(\boldsymbol\psi^{(n)})^{\mathsf T},\\[2pt]
  \mathsf M^{(n)}_2&=\boldsymbol\phi^{(n)}_1(\boldsymbol\psi^{(n)})^{\mathsf T}-\boldsymbol\phi^{(n)}(\boldsymbol\psi^{(n)}_1)^{\mathsf T},\\[2pt]
  \mathsf M^{(n)}_3&=\boldsymbol\phi^{(n)}_{11}(\boldsymbol\psi^{(n)})^{\mathsf T}-\boldsymbol\phi^{(n)}_1(\boldsymbol\psi^{(n)}_1)^{\mathsf T}+\boldsymbol\phi^{(n)}(\boldsymbol\psi^{(n)}_{11})^{\mathsf T},\\[2pt]
  \mathsf M^{(n+1)}-\mathsf M^{(n)}&=\boldsymbol\phi^{(n)}(\boldsymbol\psi^{(n+1)})^{\mathsf T},
 \end{aligned}
\end{equation}
together with the dispersion and shift relations
\begin{equation}\label{eq:waveflows}
 \begin{aligned}
  \boldsymbol\phi^{(n)}_2&=\boldsymbol\phi^{(n)}_{11},&\boldsymbol\phi^{(n)}_3&=\boldsymbol\phi^{(n)}_{111},&\boldsymbol\phi^{(n+1)}&=\boldsymbol\phi^{(n)}_1,\\
  \boldsymbol\psi^{(n)}_2&=-\boldsymbol\psi^{(n)}_{11},&\boldsymbol\psi^{(n)}_3&=\boldsymbol\psi^{(n)}_{111},&\boldsymbol\psi^{(n-1)}&=-\boldsymbol\psi^{(n)}_1.
 \end{aligned}
\end{equation}
Setting $\tau_n=\det\mathsf M^{(n)}$ and denoting Hirota's bilinear operator with respect to $x_j$ by $D_j$, one has
\begin{equation}\label{eq:hierarchy}
 (D_1^2-D_2)\tau_{n+1}\cdot\tau_n=0,\qquad (D_1^3+3D_1D_2-4D_3)\tau_{n+1}\cdot\tau_n=0.
\end{equation}
\end{lemma}
\begin{proof}
This follows from standard determinant identities in the bilinear method \cite[Chap.~3]{Hirota2004}; we record the main steps briefly. Fix $n$, and set $f=\tau_n$, $g=\tau_{n+1}$, $\mathsf M=\mathsf M^{(n)}$, $\boldsymbol\rho=\boldsymbol\phi^{(n)}$, $\boldsymbol\chi=\boldsymbol\psi^{(n+1)}$, $\boldsymbol\rho_j=\partial_1^j\boldsymbol\rho$, and $\boldsymbol\chi_j=\partial_1^j\boldsymbol\chi$. From \eqref{eq:gramflows} and \eqref{eq:waveflows}, we have $\mathsf M^{(n+1)}=\mathsf M+\boldsymbol\rho_0\boldsymbol\chi_0^{\mathsf T}$ and
\[
 \mathsf M_1=-\boldsymbol\rho_0\boldsymbol\chi_1^{\mathsf T},\qquad
 \mathsf M_2=-\boldsymbol\rho_1\boldsymbol\chi_1^{\mathsf T}+\boldsymbol\rho_0\boldsymbol\chi_2^{\mathsf T},\qquad
 \mathsf M_3=-\boldsymbol\rho_2\boldsymbol\chi_1^{\mathsf T}+\boldsymbol\rho_1\boldsymbol\chi_2^{\mathsf T}-\boldsymbol\rho_0\boldsymbol\chi_3^{\mathsf T}.
\]
For ordered tuples $I=(i_1,\ldots,i_m)$ and $J=(j_1,\ldots,j_m)$, define the signed bordered determinant $[I;J]$ by bordering $\mathsf M$ with columns $\boldsymbol\rho_{i_1},\ldots,\boldsymbol\rho_{i_m}$, rows $\boldsymbol\chi_{j_1}^{\mathsf T},\ldots,\boldsymbol\chi_{j_m}^{\mathsf T}$, and a zero $m\times m$ bottom-right block, multiplied by $(-1)^m$. Write $\Delta_{ij}=[(i);(j)]$, $\Delta=[(0,1);(0,1)]$, $\Delta_a=[(0,2);(0,1)]$, and $\Delta_b=[(0,1);(0,2)]$. Applying the rank-one perturbation identity
\[
 \det(\mathsf K+\boldsymbol\xi\boldsymbol\eta^{\mathsf T})=\det\mathsf K-\det\begin{pmatrix}\mathsf K&\boldsymbol\xi\\\boldsymbol\eta^{\mathsf T}&0\end{pmatrix}
\]
repeatedly together with \eqref{eq:waveflows} and the antisymmetry of $[I;J]$ gives $g=f+\Delta_{00}$, $f_1=-\Delta_{01}$, $g_1=\Delta_{10}$, and the combinations
\begin{align*}
 g_{11}-g_2&=2(\Delta_{11}+\Delta),& g_{11}+g_2&=2\Delta_{20},\\
 f_{11}+f_2&=-2\Delta_{11},& f_{11}-f_2&=-2\Delta_{02},\\
 g_{111}+3g_{12}-4g_3&=6(\Delta_{21}+\Delta_a-\Delta_b),&
 -f_{111}+3f_{12}+4f_3&=6\Delta_{12}.
\end{align*}
Substituting these into the left-hand sides of \eqref{eq:hierarchy} and applying the Desnanot--Jacobi determinant identity \cite{Krattenthaler1999},
\[
 f\Delta=\Delta_{00}\Delta_{11}-\Delta_{01}\Delta_{10},\qquad
 f\Delta_a=\Delta_{00}\Delta_{21}-\Delta_{01}\Delta_{20},\qquad
 f\Delta_b=\Delta_{00}\Delta_{12}-\Delta_{02}\Delta_{10},
\]
we obtain
\begin{align*}
 (D_1^2-D_2)g\cdot f&=2(f\Delta+\Delta_{10}\Delta_{01}-\Delta_{00}\Delta_{11})=0,\\
 (D_1^3+3D_1D_2-4D_3)g\cdot f&=6\bigl[f(\Delta_a-\Delta_b)-\Delta_{00}(\Delta_{21}-\Delta_{12})+\Delta_{20}\Delta_{01}-\Delta_{10}\Delta_{02}\bigr]=0. \tag*{\qed}
\end{align*}
\end{proof}

We now specify the matrix entries satisfying the hypotheses of Lemma~\ref{lem:adjacent}. Let $P,Q$ be independent spectral variables, $x_1,x_2,x_3$ be the mKP hierarchy variables, and $n\in\mathbb Z$. Since the spectral parameters used below have positive real parts, we define the kernel function on the domain $PQ(P+Q)\ne0$ by
\begin{equation}\label{eq:kernel}
 \begin{aligned}
  m^{(n)}(P,Q)&=\frac{Q}{P+Q}\left(-\frac P Q\right)^n e^{\theta(P)+\widehat\theta(Q)},\\[3pt]
  \theta(P)&=Px_1+P^2x_2+P^3x_3+\alpha(P),\\[2pt]
  \widehat\theta(Q)&=Qx_1-Q^2x_2+Q^3x_3+\widehat\alpha(Q),
 \end{aligned}
\end{equation}
where $\alpha(P)=\alpha_i(P)$ and $\widehat\alpha(Q)=\widehat\alpha_j(Q)$ are analytic phase functions for the $i$th row and $j$th column. Let $\mathscr A_i$ and $\mathscr B_j$ ($1\le i,j\le N$) be linear differential operators with constant coefficients in $P$ and $Q$, respectively, and let $p_i,q_j\in\C$ ($\re p_i,\re q_j>0$) be spectral parameters. We set
\[
\begin{aligned}
 m_{ij}^{(n)}&=\left.\mathscr A_i\mathscr B_j m^{(n)}(P,Q)\right|_{P=p_i,\,Q=q_j},\\
 \phi_i^{(n)}&=\left.\mathscr A_i\bigl(P^n e^{\theta(P)}\bigr)\right|_{P=p_i},\\
 \psi_j^{(n)}&=\left.\mathscr B_j\bigl(Q(-Q)^{-n}e^{\widehat\theta(Q)}\bigr)\right|_{Q=q_j}.
\end{aligned}
\]
Differentiating $m^{(n)}(P,Q)$ with respect to $x_1,x_2,x_3$ cancels the denominator $P+Q$, while
\[
 m^{(n+1)}(P,Q)-m^{(n)}(P,Q)=P^ne^{\theta(P)}\cdot Q(-Q)^{-n-1}e^{\widehat\theta(Q)}.
\]
Since $\mathscr A_i$ and $\mathscr B_j$ commute with $\partial_{x_1},\partial_{x_2},\partial_{x_3}$ and act separately on the $P$- and $Q$-factors, evaluating at $(P,Q)=(p_i,q_j)$ yields \eqref{eq:gramflows}--\eqref{eq:waveflows}, and Lemma~\ref{lem:adjacent} implies \eqref{eq:hierarchy}.

Under the variable transformation
\begin{equation}\label{eq:coord}
 x_1=x,\qquad x_2=-\ii y,\qquad x_3=-4t,
\end{equation}
we have $D_1=D_x$, $D_2=\ii D_y$, and $D_3=-D_t/4$, so that \eqref{eq:hierarchy} for $n=0$ becomes
\[
 (D_x^2-\ii D_y)\tau_1\cdot\tau_0=0,\qquad (D_x^3+3\ii D_xD_y+D_t)\tau_1\cdot\tau_0=0.
\]
By the identity
\[
 D_x^mD_y^nD_t^k(e^\Theta g)\cdot(e^\Theta f)=e^{2\Theta}D_x^mD_y^nD_t^k g\cdot f
\]
for any affine function $\Theta$ of $(x,y,t)$, whenever $\tau_0=c\,e^\Theta f$ and $\tau_1=\tilde c\,e^\Theta f^*$ with nonzero constants $c,\tilde c$ satisfy \eqref{eq:hierarchy} for $n=0$, dividing out $c\tilde c\,e^{2\Theta}$ and multiplying the first equation by $\ii$ shows that $f$ satisfies \eqref{eq:bilinear}.

\subsection{Determinants for standard lump solutions}\label{sec:standard}

We rewrite $f_{{\rm std},M}$ as an $M\times M$ Gram determinant in integral form (Lemmas~\ref{lem:cycles}--\ref{lem:bridge}) and thereby prove Theorem~\ref{th:construction}(i).

\begin{lemma}\label{lem:cycles}
In the Leibniz expansion of $f_{{\rm std},M}=\det\mathsf A$ in \eqref{eq:standard}, all terms containing a cycle of length at least $3$ cancel. Consequently, $\det\mathsf A$ expands solely over involutions on $\{1,\ldots,2M\}$: each fixed point $j$ contributes $\mathsf A_{jj}$, and each transposition $(j,k)$ contributes $-4/(\lambda_j-\lambda_k)^2$.
\end{lemma}
\begin{proof}
Decompose each permutation in the Leibniz expansion of $\det\mathsf A$ into disjoint cycles. On any $m$-element subset of indices (say $\{1,\ldots,m\}$) with $m\ge3$, all $m$-cycles have the same sign and numerator $(2\ii)^m$. Every $m$-cycle on $\{1,\ldots,m\}$ is uniquely obtained from an $(m-1)$-cycle on $\{1,\ldots,m-1\}$ by inserting $m$ into a directed edge $u\to v$, which replaces $(\lambda_u-\lambda_v)^{-1}$ in the cycle product by
\[
 \frac{1}{(\lambda_u-\lambda_m)(\lambda_m-\lambda_v)}
 =\frac{1}{\lambda_u-\lambda_v}\left(\frac{1}{\lambda_u-\lambda_m}-\frac{1}{\lambda_v-\lambda_m}\right).
\]
Summing over all $m-1$ edges $u\to v$ of a fixed $(m-1)$-cycle gives zero because each vertex appears once as $u$ and once as $v$. Thus all terms containing a cycle of length $m\ge3$ cancel, leaving only involutions: each fixed point $j$ contributes $\mathsf A_{jj}$, and each transposition $(j,k)$ has sign $-1$ and contributes $-4/(\lambda_j-\lambda_k)^2$. \qed
\end{proof}

\begin{lemma}\label{lem:bridge}
Set $\mu_j=\ii\lambda_j^*/2$, and define
\[
 \begin{aligned}
  \kappa_j&=-\sum_{\ell\ne j}\frac1{\mu_j-\mu_\ell}
        +\sum_{\ell=1}^M\frac1{\mu_j+\mu_\ell^*}
        -\frac1{2\mu_j}+\gamma_j^*,\\[4pt]
  w_j&=x-2\ii\mu_jy-12\mu_j^2t+\kappa_j.
 \end{aligned}
\]
For $s\le0$, let $\phi_j(s)=e^{\mu_js}(s+w_j)$ and define the $M\times M$ matrix $\mathsf M$ by
\[
 \mathsf M_{jk}=\int_{-\infty}^0\phi_j(s)(\partial_s\phi_k)^*(s)\dd s .
\]
Then the $\tau$-function $f_{{\rm std},M}$ and the Gram determinant $\det\mathsf M$ satisfy
\begin{equation}\label{eq:bridge}
 \det\mathsf M=\mathcal C_M f_{{\rm std},M},\qquad\mathcal C_M=\det\left[\frac{\mu_k^*}{\mu_j+\mu_k^*}\right]_{1\le j,k\le M}\ne0.
\end{equation}
\end{lemma}
\begin{proof}
Since $\re\mu_j=b_j/2>0$, these integrals converge absolutely. Substituting $(\partial_s\phi_k)^*(s)=e^{\mu_k^*s}[1+\mu_k^*(s+w_k^*)]$ and using $\int_{-\infty}^0 s^m e^{cs}\dd s=(-1)^m m!\,c^{-m-1}$ ($m=0,1,2$) with $c=c_{jk}=\mu_j+\mu_k^*$ gives
\[
 \mathsf M_{jk}=\frac{\mu_k^*w_jw_k^*}{c_{jk}}+\frac{\mu_jw_j-\mu_k^*w_k^*}{c_{jk}^2}+\frac{\mu_k^*-\mu_j}{c_{jk}^3}.
\]
Next, introduce independent row and column spectral variables $\boldsymbol P=(P_1,\ldots,P_M)$ and $\boldsymbol Q=(Q_1,\ldots,Q_M)$, and take the linear phases $\alpha_j(P_j)=\kappa_jP_j$ and $\widehat\alpha_k(Q_k)=\kappa_k^*Q_k$ in \eqref{eq:kernel}. Under \eqref{eq:coord}, the row and column exponential factors are
\[
 \begin{aligned}
  E_j(P_j)&=\exp(P_jx-\ii P_j^2y-4P_j^3t+\kappa_jP_j),\\[2pt]
  \widehat E_k(Q_k)&=\exp(Q_kx+\ii Q_k^2y-4Q_k^3t+\kappa_k^*Q_k).
 \end{aligned}
\]
At $P_j=\mu_j$ and $Q_k=\mu_k^*$, the first logarithmic derivatives of $E_j$ and $\widehat E_k$ are $w_j$ and $w_k^*$, while the partial derivatives $\partial_{P_j}$, $\partial_{Q_k}$, and $\partial_{P_j}\partial_{Q_k}$ of $Q_k/(P_j+Q_k)$ are $-\mu_k^*/c_{jk}^2$, $\mu_j/c_{jk}^2$, and $(\mu_k^*-\mu_j)/c_{jk}^3$. By the Leibniz rule,
\[
 \left.\partial_{P_j}\partial_{Q_k}\left(\frac{Q_k}{P_j+Q_k}E_j(P_j)\widehat E_k(Q_k)\right)\right|_{P_j=\mu_j,\,Q_k=\mu_k^*}
 =E_j(\mu_j)\widehat E_k(\mu_k^*)\mathsf M_{jk}.
\]

Let $\mathcal C(\boldsymbol P,\boldsymbol Q)=\det[Q_k/(P_j+Q_k)]$ denote the Cauchy determinant. By the Cauchy determinant formula \cite{Krattenthaler1999},
\[
 \mathcal C(\boldsymbol P,\boldsymbol Q)=\frac{\prod_kQ_k\prod_{j<k}(P_k-P_j)(Q_k-Q_j)}
 {\prod_{j,k}(P_j+Q_k)}.
\]
For $\mathcal I=\{P_1,\ldots,P_M,Q_1,\ldots,Q_M\}$, set
\[
 F(\boldsymbol P,\boldsymbol Q)=\mathcal C(\boldsymbol P,\boldsymbol Q)\prod_{j=1}^M E_j(P_j)\widehat E_j(Q_j),\qquad
 \partial_{\mathcal I}=\prod_{j=1}^M\partial_{P_j}\partial_{Q_j}.
\] Since $P_j$ appears only in the $j$th row and $Q_k$ appears only in the $k$th column of the determinant, applying $\partial_{\mathcal I}$ to $F$ differentiates the $(j,k)$ entry by $\partial_{P_j}\partial_{Q_k}$, giving
\[
 \left.\partial_{\mathcal I}F\right|_{\boldsymbol P=\boldsymbol\mu,\ \boldsymbol Q=\boldsymbol\mu^*}
 =\left(\prod_{j=1}^M E_j(\mu_j)\widehat E_j(\mu_j^*)\right)\det\mathsf M .
\]
We now evaluate the left-hand side via logarithmic derivatives. Since each factor in the product defining $F$ depends on at most two spectral variables, all mixed partial derivatives of $\log F$ with respect to three or more distinct spectral variables vanish. Expanding $\partial_{\mathcal I}F/F$ over set partitions of $\mathcal I$ therefore involves only partitions $\mathcal P$ into blocks of size one or two:
\[
 \frac{\partial_{\mathcal I}F}{F}
  =\sum_{\mathcal P}\prod_{I\in\mathcal P}\partial_I\log F.
\]
For singleton blocks, using
\[
 \partial_{P_j}\log\mathcal C=\sum_{\ell\ne j}\frac1{P_j-P_\ell}-\sum_k\frac1{P_j+Q_k},\qquad
 \partial_{Q_j}\log\mathcal C=\frac1{Q_j}+\sum_{\ell\ne j}\frac1{Q_j-Q_\ell}-\sum_k\frac1{P_k+Q_j},
\]
together with the definition of $\kappa_j$ and the relations $\lambda_{M+j}=\lambda_j^*=-2\ii\mu_j$ and $\lambda_j=2\ii\mu_j^*$, we find
\begin{align*}
 \left.\partial_{P_j}\log F\right|_{\boldsymbol P=\boldsymbol\mu,\,\boldsymbol Q=\boldsymbol\mu^*}
 &=w_j+\left.\partial_{P_j}\log\mathcal C\right|_{\boldsymbol P=\boldsymbol\mu,\,\boldsymbol Q=\boldsymbol\mu^*}
 =x+\lambda_{M+j}y+3\lambda_{M+j}^2t+\frac{\ii}{\lambda_{M+j}}+\gamma_{M+j},\\
 \left.\partial_{Q_j}\log F\right|_{\boldsymbol P=\boldsymbol\mu,\,\boldsymbol Q=\boldsymbol\mu^*}
 &=w_j^*+\left.\partial_{Q_j}\log\mathcal C\right|_{\boldsymbol P=\boldsymbol\mu,\,\boldsymbol Q=\boldsymbol\mu^*}
 =x+\lambda_jy+3\lambda_j^2t+\frac{\ii}{\lambda_j}+\gamma_j,
\end{align*}
which are the $(M+j)$th and $j$th diagonal entries of $\mathsf A$, respectively. For two-element blocks, the mixed second derivatives of $\log F$ reduce to those of $\log\mathcal C$:
\[
 \partial_{P_j}\partial_{P_k}\log\mathcal C=\frac1{(P_j-P_k)^2},\qquad
 \partial_{Q_j}\partial_{Q_k}\log\mathcal C=\frac1{(Q_j-Q_k)^2},\qquad
 \partial_{P_j}\partial_{Q_k}\log\mathcal C=\frac1{(P_j+Q_k)^2}.
\]
At $P_j=\ii\lambda_{M+j}/2$ and $Q_k=-\ii\lambda_k/2$, these three derivatives equal $-4/(\lambda_{M+j}-\lambda_{M+k})^2$, $-4/(\lambda_j-\lambda_k)^2$, and $-4/(\lambda_{M+j}-\lambda_k)^2$, matching the transposition factors in Lemma~\ref{lem:cycles}.

By Lemma~\ref{lem:cycles}, the value of $\partial_{\mathcal I}F/F$ at $(\boldsymbol\mu,\boldsymbol\mu^*)$ is therefore $\det\mathsf A=f_{{\rm std},M}$. Equating the two expressions for $\partial_{\mathcal I}F$ and dividing out the nonzero exponential factor gives \eqref{eq:bridge}, with
\[
 \mathcal C_M=\mathcal C(\boldsymbol\mu,\boldsymbol\mu^*)=
 \frac{\left(\prod_j\mu_j^*\right)\prod_{j<k}|\mu_k-\mu_j|^2}
 {\left(\prod_j b_j\right)\prod_{j<k}|\mu_j+\mu_k^*|^2}\ne0. \tag*{\qed}
\]
\end{proof}

\begin{theopargself}
\begin{proof}[of Theorem~\ref{th:construction}\textup{(i)}]
In the kernel \eqref{eq:kernel}, choose the linear phases and spectral parameters of Lemma~\ref{lem:bridge}, apply $\partial_{P_j}\partial_{Q_k}$ to the $(j,k)$ entry, evaluate at $(P_j,Q_k)=(\mu_j,\mu_k^*)$, and impose \eqref{eq:coord}. Denote the resulting matrices for $n=0,1$ by $\mathsf M^{(0)}$ and $\mathsf M^{(1)}$. Since $(E_j(P))^*=\widehat E_j(P^*)$ for real $(x,y,t)$, taking the conjugate transpose interchanges the roles of $P_j$ and $Q_k$ as well as the evaluation points $\mu_j$ and $\mu_k^*$. Since the prefactor $Q_k(-P_j/Q_k)^n$ in the kernel is $Q_k$ for $n=0$ and $-P_j$ for $n=1$, we obtain $\mathsf M^{(1)}=-(\mathsf M^{(0)})^\dagger$, so that $\tau_0^*=(-1)^M\tau_1$.

By the proof of Lemma~\ref{lem:bridge},
\[
 (\mathsf M^{(0)})_{jk}=E_j(\mu_j)\widehat E_k(\mu_k^*)\,\mathsf M_{jk},\qquad e^\Theta=\prod_{j=1}^M E_j(\mu_j)\widehat E_j(\mu_j^*),
\]
where $\Theta$ is a real affine function of $(x,y,t)$. Together with \eqref{eq:bridge}, this gives $\tau_0=e^\Theta\mathcal C_M f_{{\rm std},M}$ and $\tau_1=(-1)^M e^\Theta\mathcal C_M^* f_{{\rm std},M}^*$. Substituting these into \eqref{eq:hierarchy} and canceling $(-1)^M e^{2\Theta}|\mathcal C_M|^2$ shows that $f_{{\rm std},M}$ satisfies \eqref{eq:bilinear}. \qed
\end{proof}
\end{theopargself}

As a special case, consider the fundamental lump solution for $M=1$ and $\lambda=a+\ii b$ \cite{QWY}. In terms of the co-moving coordinates
\[
 X=x-3(a^2+b^2)t,\qquad Y=y+6at,\qquad Z=X+aY,
\]
the $\tau$-function \eqref{eq:standard} for $\gamma_1=0$ takes the form
\begin{equation}\label{eq:fundamental}
 f_0(Z,Y)=Z^2+b^2Y^2+\frac{a^2}{b^2(a^2+b^2)}+\frac{2\ii(aZ-b^2Y)}{a^2+b^2},
\end{equation}
while a general phase $\gamma_1\in\C$ simply translates $(Z,Y)$ to $(Z+\re\gamma_1,Y+\im\gamma_1/b)$. When $a\ne0$, $\re f_0(Z,Y)>0$ and $v$ attains the extremum $8b^2/a$ at $(Z,Y)=(0,0)$; when $a=0$, $f_0(Z,Y)=Z^2+b^2Y^2-2\ii Y$ vanishes at the origin. Thus, $a\ne0$ is necessary and sufficient for the regularity of the fundamental lump. This confirms the sharpness of Theorem~\ref{th:regularity}(i) for $M=1$ and of the threshold $|a|>b\cot(\pi/2)=0$ in \eqref{eq:cotbound} for $\Lambda=(1)$.

\subsection{Determinants for higher-order lump solutions}\label{sec:higher}

We now prove Theorem~\ref{th:construction}(ii) by Taylor expansion at a single conjugate pair of spectral parameters, and then express $f_\Lambda$ as a Gram determinant in integral form.

\begin{theopargself}
\begin{proof}[of Theorem~\ref{th:construction}\textup{(ii)}]
For the single conjugate pair of spectral parameters in Theorem~\ref{th:construction}(ii), we have $\mu=(b+\ii a)/2$, $q=\mu^*$, and $\mu+q=b$. In the kernel \eqref{eq:kernel}, choose an analytic phase function $\alpha(P)$ near $\mu$ and set $\widehat\alpha(Q)=(\alpha(Q^*))^*$, with Taylor expansions
\[
 \alpha(\mu+\epsilon)=\sum_{k\ge0}\zeta_k\epsilon^k,\qquad
 \widehat\alpha(q+\delta)=\sum_{k\ge0}\zeta_k^*\delta^k,
\]
where $\zeta_0=0$, and any given $\zeta_1,\ldots,\zeta_d\in\C$ can be realized by a polynomial $\alpha(P)$ of degree at most $d$.

For $1\le i,j\le N$, define the differential operators with respect to $P,Q$ and the determinants
\[
 \begin{aligned}
  \mathscr A_i&=\frac{1}{n_i!}\partial_P^{n_i},\qquad
  \mathscr B_j=\frac{1}{n_j!}\partial_Q^{n_j},\\[3pt]
  \tau_n&=\left.\det\bigl[\mathscr A_i\mathscr B_j m^{(n)}(P,Q)\bigr]_{i,j=1}^N\right|_{P=\mu,\,Q=q}.
 \end{aligned}
\]
With $P=\mu+\epsilon$ and $Q=q+\delta$, the operators $\mathscr A_i$ and $\mathscr B_j$ extract the coefficients of $\epsilon^{n_i}$ and $\delta^{n_j}$ in the Taylor expansion at $(\epsilon,\delta)=(0,0)$. Denote the resulting matrices for $n=0,1$ by $\mathsf M^{(0)}$ and $\mathsf M^{(1)}$. Since $\theta(P)^*=\widehat\theta(P^*)$ under \eqref{eq:coord}, taking the complex conjugate of the $(j,i)$ entry of $\mathsf M^{(0)}$ interchanges the differentiation orders $n_j$ and $n_i$ with respect to $P$ and $Q$ as well as the spectral parameters $\mu$ and $q$. Since the prefactor $Q(-P/Q)^n$ in the kernel is $Q$ for $n=0$ and $-P$ for $n=1$, we again have $\mathsf M^{(1)}=-(\mathsf M^{(0)})^\dagger$, and thus $\tau_0^*=(-1)^N\tau_1$.

Factoring out $e^\Theta$, where $\Theta=N[\theta(\mu)+\widehat\theta(q)]$ is a real linear function of $(x,y,t)$, and expanding $\theta(\mu+\epsilon)-\theta(\mu)$ in powers of $\epsilon$ produces the variables $\boldsymbol s=(s_1,s_2,\ldots)$ in \eqref{eq:physicalschur}. Once the common exponential factor is removed, the $(i,j)$ entry of $\mathsf M^{(0)}$ is the coefficient of $\epsilon^{n_i}\delta^{n_j}$ in
\[
 \frac{q+\delta}{b+\epsilon+\delta}\exp\!\left(\sum_{m\ge1}s_m\epsilon^m\right)\exp\!\left(\sum_{m\ge1}s_m^*\delta^m\right).
\]
By \eqref{eq:schurseries}, the definition of $\mathsf H_{k\ell}$ in \eqref{eq:highertau}, and the Cauchy product rule, this coefficient is given by
\[
 \sum_{k=0}^{n_i}\sum_{\ell=0}^{n_j}S_{n_i-k}(\boldsymbol s)\,\mathsf H_{k\ell}\,S_{n_j-\ell}^*(\boldsymbol s)=(\mathsf P\mathsf H\mathsf P^\dagger)_{ij},
\]
where $S_{n_i-k}=0$ for $k>n_i$ and $S_{n_j-\ell}=0$ for $\ell>n_j$. Taking determinants gives $\tau_0=e^\Theta f_\Lambda$ and $\tau_1=(-1)^N e^\Theta f_\Lambda^*$. Substituting these into \eqref{eq:hierarchy} and canceling $(-1)^N e^{2\Theta}$ proves that $f_\Lambda$ satisfies \eqref{eq:bilinear}. \qed
\end{proof}
\end{theopargself}

The $\tau$-function $f_\Lambda$ also admits a Gram determinant representation in integral form. For $\re(b+\epsilon+\delta)>0$, we have
\[
 \frac{q+\delta}{b+\epsilon+\delta}=\int_{-\infty}^0(q+\delta)e^{(b+\epsilon+\delta)s}\dd s .
\]
Since $b>0$, the integrand and all its partial derivatives with respect to $(\epsilon,\delta)$ decay exponentially as $s\to-\infty$, uniformly for $(\epsilon,\delta)$ in a small neighborhood of $(0,0)$, so extracting Taylor coefficients commutes with integration. Multiplying $e^{(\mu+\epsilon)s}$ into $\exp(\sum_{m\ge1}s_m\epsilon^m)$ shifts $s_1$ to $s_1+s$; the coefficient of $\epsilon^{n_i}$ is the exponential polynomial
\[
 \phi_i(s)=e^{\mu s}S_{n_i}(s_1+s,s_2,s_3,\ldots)=e^{\mu s}\sum_{k=0}^{n_i}S_{n_i-k}(\boldsymbol s)\frac{s^k}{k!},
\]
whose polynomial part in $s$ has degree $n_i$ and leading coefficient $1/n_i!$. Likewise, since $(q+\delta)e^{(q+\delta)s}=\mathcal D\bigl(e^{(q+\delta)s}\bigr)$ with $\mathcal D=\partial_s$ and $q=\mu^*$, extracting the coefficient of $\delta^{n_j}$ from the conjugate factor yields $(\mathcal D\phi_j)^*(s)$. This gives the integral representation
\begin{equation}\label{eq:highergram}
 (\mathsf P\mathsf H\mathsf P^\dagger)_{ij}=\int_{-\infty}^0\phi_i(s)(\mathcal D\phi_j)^*(s)\dd s .
\end{equation}

\begin{remark}
By \eqref{eq:bridge} and \eqref{eq:highergram}, the $\tau$-functions of both standard and higher-order lump solutions take the form $\det\bigl[\int_{-\infty}^0\phi_i(s)(\mathcal D\phi_j)^*(s)\dd s\bigr]$ up to a nonzero constant factor. In Sect.~\ref{sec:regularity}, we use this common Gram determinant structure to prove the regularity of both classes of solutions via quadratic form estimates for $\mathcal D$.
\end{remark}

\section{Regularity and spatial decay}\label{sec:regularity}

\subsection{Operator estimates and the spectrum of \texorpdfstring{$\mathsf S_n$}{S\_n}}\label{sec:tools}

Let $\langle g,h\rangle=\int_{-\infty}^0g^*(s)h(s)\dd s$ denote the $L^2$ inner product on $(-\infty,0]$ and $\mathcal D=\partial_s$. Suppose $\mathcal W\subset L^2(-\infty,0]$ is a finite-dimensional $\mathcal D$-invariant subspace of exponential polynomials that vanish as $s\to-\infty$. With the evaluation functional $\mathcal E g=g(0)$, integration by parts yields, for all $g,h\in\mathcal W$,
\[
 \langle g,\mathcal D h\rangle+\langle\mathcal Dg,h\rangle
 =\bigl[g^*(s)h(s)\bigr]_{s=-\infty}^{s=0}
 =g(0)^*h(0)=\langle\mathcal E g,\mathcal E h\rangle_{\C}.
\]
Denoting the adjoint of $\mathcal D$ on $\mathcal W$ by $\mathcal D^\dagger$, we obtain
\begin{equation}\label{eq:boundary}
 \mathcal D+\mathcal D^\dagger=\mathcal E^\dagger\mathcal E ,
\end{equation}
Setting $\im\mathcal D=(\mathcal D-\mathcal D^\dagger)/(2\ii)$ and taking $h=g$ in \eqref{eq:boundary} gives $\langle g,(\im\mathcal D)g\rangle=\im\langle g,\mathcal Dg\rangle$ and $2\re\langle g,\mathcal D g\rangle=|g(0)|^2$.

\Needspace{6\baselineskip}
\begin{lemma}\label{lem:endpoint}
Suppose that there exists $\sigma\in\{1,-1\}$ such that $\sigma\langle g,(\im\mathcal D)g\rangle>0$ for all nonzero $g\in\ker\mathcal E$. For linearly independent functions $\phi_1,\ldots,\phi_N\in\mathcal W$, define the $N\times N$ matrices
\[
 \mathsf M_{ij}=\int_{-\infty}^0\phi_i(s)(\mathcal D\phi_j)^*(s)\dd s,\qquad
 \mathsf B_{ij}=\int_{-\infty}^0\phi_i(s)\phi_j^*(s)\dd s .
\]
Then the Gram matrix $\mathsf B$ is positive definite, and there exists a constant $c_{\mathcal D}>0$, depending only on $\mathcal D$ and $\mathcal W$, such that $|\det\mathsf M|\ge c_{\mathcal D}^N\det\mathsf B$.
\end{lemma}
\begin{proof}
For any $z\in\C^N$, set $g=\sum_{i=1}^Nz_i^*\phi_i$. Since $\mathsf B_{ij}=\langle\phi_j,\phi_i\rangle$ and $\mathsf M_{ij}=\langle\mathcal D\phi_j,\phi_i\rangle$, we have
\[
 z^\dagger \mathsf B z=\|g\|^2,\qquad
 z^\dagger \mathsf M z=\langle g,\mathcal D g\rangle^*.
\]
Because $\phi_1,\ldots,\phi_N$ are linearly independent, $\mathsf B$ is positive definite. If $\langle g,\mathcal D g\rangle=0$, taking the real part and using \eqref{eq:boundary} gives $|g(0)|^2=0$, so $g\in\ker\mathcal E$; taking the imaginary part and using the hypothesis on $\ker\mathcal E$ then gives $g=0$. Hence, $|\langle g,\mathcal Dg\rangle|$ is continuous and strictly positive on the compact unit sphere $\|g\|=1$ in $\mathcal W$, attaining a minimum
\[
 c_{\mathcal D}=\min_{\substack{g\in\mathcal W\\ \|g\|=1}}|\langle g,\mathcal D g\rangle|>0.
\]
By homogeneity, $|z^\dagger\mathsf Mz|\ge c_{\mathcal D}z^\dagger\mathsf Bz$.

Setting $\widetilde{\mathsf M}=\mathsf B^{-1/2}\mathsf M\mathsf B^{-1/2}$ and replacing $z$ by $\mathsf B^{-1/2}z$ gives
\[
 |z^\dagger\widetilde{\mathsf M}z|\ge c_{\mathcal D}\|z\|^2 .
\]
By the Cauchy--Schwarz inequality, $\|\widetilde{\mathsf M}z\|\ge c_{\mathcal D}\|z\|$, so the smallest singular value of $\widetilde{\mathsf M}$ is at least $c_{\mathcal D}$, and $|\det\widetilde{\mathsf M}|\ge c_{\mathcal D}^N$. Since $\det\mathsf B>0$, we obtain $|\det\mathsf M|=|\det\widetilde{\mathsf M}|\det\mathsf B\ge c_{\mathcal D}^N\det\mathsf B$. \qed
\end{proof}

Let $\mathsf S_n$ be the $n\times n$ Hermitian Toeplitz matrix with zero main diagonal, strictly upper-triangular entries $-\ii$, and strictly lower-triangular entries $\ii$, and let $\mathbf1=(1,\ldots,1)^{\mathsf T}\in\C^n$. The next lemma gives the eigenvalues of $\mathsf S_n$ and of its compression to $\mathbf1^\perp$.

\begin{lemma}\label{lem:signspectrum}
For an integer $n\ge2$, the eigenvalues of $\mathsf S_n$ are
\[
 \left\{\cot\frac{(2k-1)\pi}{2n}:1\le k\le n\right\}.
\]
If $\Pi$ denotes the orthogonal projection onto $\mathbf1^\perp$, then the eigenvalues of the compression $\Pi\mathsf S_n|_{\mathbf1^\perp}$ of $\mathsf S_n$ to $\mathbf1^\perp$ are
\[
 \left\{\cot\frac{\pi k}{n}:1\le k<n\right\}.
\]
\end{lemma}
\begin{proof}
Index the vector components by $0,1,\ldots,n-1$. For $r\ne1$, set $\boldsymbol r=(1,r,\ldots,r^{n-1})^{\mathsf T}$. Summing the geometric series below and above the main diagonal gives
\begin{equation}\label{eq:geometric}
 (\mathsf S_n\boldsymbol r)_j=\ii\sum_{\ell<j}r^\ell-\ii\sum_{\ell>j}r^\ell
       =\frac{\ii(1+r^n)}{1-r}-\ii\frac{1+r}{1-r}r^j .
\end{equation}
When $r$ is one of the $n$ distinct roots of $r^n=-1$, the first term on the right-hand side of \eqref{eq:geometric} vanishes; since these vectors form the columns of a nonsingular Vandermonde matrix, they are linearly independent and form an eigenbasis of $\C^n$. For $\Pi\mathsf S_n|_{\mathbf1^\perp}$, take the $n-1$ roots of $r^n=1$ with $r\ne1$. Since $\sum_{j=0}^{n-1}r^j=0$, the vectors $\boldsymbol r$ lie in $\mathbf1^\perp$ and are linearly independent, spanning $\mathbf1^\perp$. The $j$-independent constant term in \eqref{eq:geometric} is annihilated by $\Pi$, so these vectors are eigenvectors of $\Pi\mathsf S_n|_{\mathbf1^\perp}$. For $r=e^{\ii\theta}$, the corresponding eigenvalue in both cases is $-\ii(1+r)/(1-r)=\cot(\theta/2)$. Substituting $\theta=(2k-1)\pi/n$ ($1\le k\le n$) and $\theta=2\pi k/n$ ($1\le k<n$) gives the stated spectra. \qed
\end{proof}

By the symmetry of both spectra about the origin, the smallest eigenvalue of $\pm\mathsf S_n$ on $\C^n$ is $-\cot(\pi/(2n))$, while the smallest eigenvalue of $\Pi(\pm\mathsf S_n)|_{\mathbf1^\perp}$ is $-\cot(\pi/n)$.

\subsection{Spatial decay of lump solutions}\label{sec:decay}

Suppose that a $\tau$-function $f$ satisfying \eqref{eq:bilinear} has no zeros in $\R^3$ and is a polynomial in $(x,y)$ with smooth $t$-dependent coefficients. Let $f_{\rm lead}$ denote its highest-degree homogeneous part in $(x,y)$, and assume that $f_{\rm lead}$ is independent of $t$, has constant argument, and vanishes nowhere on the unit circle. We show that $v$ and all its spatial partial derivatives decay algebraically as $R\to\infty$ uniformly on compact time intervals, and that $v$ satisfies Eq.~\eqref{eq:mkp}.

Let $L$ be the degree of $f_{\rm lead}$ in $(x,y)$. For a multi-index $\beta=(\beta_x,\beta_y)$ of nonnegative integers, write $\partial_{x,y}^\beta=\partial_x^{\beta_x}\partial_y^{\beta_y}$ and $|\beta|=\beta_x+\beta_y$. On any compact time interval and for $R\ge1$, the remainder $f-f_{\rm lead}$ satisfies $\partial_{x,y}^\beta(f-f_{\rm lead})=O(R^{L-1-|\beta|})$, while $\partial_{x,y}^\beta(1/f_{\rm lead})$ is homogeneous of degree $-L-|\beta|$ and bounded on the unit circle, hence $O(R^{-L-|\beta|})$. Setting
\[
 r=\frac{f-f_{\rm lead}}{f_{\rm lead}},\qquad f=f_{\rm lead}(1+r),
\]
we obtain $\partial_{x,y}^\beta r=O(R^{-1-|\beta|})$ from the Leibniz rule. For $R$ sufficiently large, uniformly on the compact time interval, we have $|r|\le 1/2$ and $|1+r|\ge 1/2$. By induction on $|\beta|$ via $(1+r)(1+r)^{-1}=1$, we obtain $\partial_{x,y}^\beta(1+r)^{-1}=O(R^{-1-|\beta|})$ for all $|\beta|>0$. The Leibniz rule then yields $\partial_{x,y}^\beta[r_x/(1+r)]=O(R^{-2-|\beta|})$ and $\partial_{x,y}^\beta[r_y/(1+r)]=O(R^{-2-|\beta|})$. Since $f_{\rm lead}$ has constant argument, both $(f_{\rm lead})_x/f_{\rm lead}$ and $(f_{\rm lead})_y/f_{\rm lead}$ are real-valued, so that $v$ and the auxiliary function $V=4\im(f_y/f)$ reduce to
\[
 v=4\im\frac{r_x}{1+r},\qquad V=4\im\frac{r_y}{1+r}.
\]
Thus, for every multi-index $\beta$, we have, uniformly on compact time intervals,
\[
 \partial_{x,y}^\beta v=O(R^{-2-|\beta|}),\qquad\partial_{x,y}^\beta V=O(R^{-2-|\beta|}).
\]

To derive the coupled local equations for $v$ and $V$ from \eqref{eq:bilinear}, choose a global smooth branch $\ell_f$ of $\log(f^*/f)$ on $\R^3$ and set $\Omega=\log|f|^2$. By the standard logarithmic-derivative identities \cite[Chap.~1]{Hirota2004}, dividing \eqref{eq:bilinear} by $|f|^2$ gives
\[
 \begin{aligned}
  \Omega_{xx}+\ell_{f,x}^2-\ii\ell_{f,y}&=0,\\[2pt]
  \ell_{f,t}+\ell_{f,xxx}+3\ell_{f,x}\Omega_{xx}+\ell_{f,x}^3+3\ii(\Omega_{xy}+\ell_{f,x}\ell_{f,y})&=0.
 \end{aligned}
\]
Eliminate $\Omega_{xx}$ using the first equation and differentiate the second with respect to $x$. Substituting $\Omega_{xxy}=-2\ell_{f,x}\ell_{f,xy}+\ii\ell_{f,yy}$, multiplying by $2\ii$, and setting $v=2\ii\ell_{f,x}$ and $V=2\ii\ell_{f,y}$ yields the coupled local system
\[
 \begin{aligned}
  v_t+v_{xxx}+\frac32v^2v_x+3v_xV-3V_y&=0,\\[2pt]
  V_x&=v_y.
 \end{aligned}
\]

These decay estimates imply that $V=O(|x|^{-2})$, $V_y=O(|x|^{-3})$, $v_y=O(|x|^{-3})$, and $v_{yy}=O(|x|^{-4})$ as $x\to-\infty$, uniformly in $y\in\R$ and on compact intervals in $t$. Hence $V,V_y\to0$ as $x\to-\infty$, and $v_y,v_{yy}$ are integrable on $(-\infty,x]$. Integrating $V_x=v_y$ and $(V_y)_x=V_{yx}=v_{yy}$ from $-\infty$ to $x$ gives
\[
 V(x,y,t)=\int_{-\infty}^xv_y(x',y,t)\dd x',\qquad
 V_y(x,y,t)=\int_{-\infty}^xv_{yy}(x',y,t)\dd x',
\]
and substituting these into the first equation of the coupled local system shows that $v=4\im(f_x/f)$ satisfies Eq.~\eqref{eq:mkp}.

\subsection{Proof of Theorem~\ref{th:regularity}}\label{sec:regproofs}

\begin{theopargself}
\begin{proof}[of Theorem~\ref{th:regularity}\textup{(i)}]
Let $\mathbb P_k(s)$ denote the space of polynomials in $s$ of degree at most $k$. Each two-dimensional subspace $e^{\mu_js}\mathbb P_1(s)$ is $\mathcal D$-invariant. Since $\mu_1,\ldots,\mu_M$ are distinct, the operator $\prod_{k\ne j}(\mathcal D-\mu_k)^2$ annihilates $e^{\mu_ks}\mathbb P_1(s)$ for all $k\ne j$ and is invertible on $e^{\mu_js}\mathbb P_1(s)$; the direct sum of these $M$ subspaces therefore forms a $2M$-dimensional $\mathcal D$-invariant space $\mathcal V$. Because the functions $\phi_j(s)=e^{\mu_js}(s+w_j)$ are nonzero and belong to distinct direct summands, $\phi_1,\ldots,\phi_M$ are linearly independent in $\mathcal V$ for all $(x,y,t)\in\R^3$ and any fixed phase parameters.

Order the natural basis of $\mathcal V$ as $(e^{\mu_1s},se^{\mu_1s},\ldots,e^{\mu_Ms},se^{\mu_Ms})$, and set
\[
 \begin{aligned}
  (\widehat a_{2j-1},\nu_{2j-1})&=(\widehat a_{2j},\nu_{2j})=\bigl(a_j,\sqrt{b_j}\bigr)\qquad(1\le j\le M),\\[2pt]
  \widehat\mu_\ell&=\frac12(\nu_\ell^2+\ii\widehat a_\ell)\qquad(1\le\ell\le2M).
 \end{aligned}
\]
Since $\mathcal D e^{\mu_js}=\mu_je^{\mu_js}$ and $\mathcal D(se^{\mu_js})=\mu_jse^{\mu_js}+e^{\mu_js}$, the span of the first $\ell$ basis vectors is $\mathcal D$-invariant for each $\ell$. Applying Gram--Schmidt orthonormalization in this order preserves the successive spans and produces an orthonormal basis $\{u_\ell\}_{\ell=1}^{2M}$ in which $\mathcal D$ is represented by an upper-triangular matrix $\mathsf D=(\mathsf D_{\ell k})$ ($\mathsf D_{\ell k}=\langle u_\ell,\mathcal D u_k\rangle$) with diagonal entries $\widehat\mu_\ell$.

In this orthonormal basis, the evaluation functional $\mathcal E$ is represented by the row vector $\bigl(u_1(0),\ldots,u_{2M}(0)\bigr)$. From \eqref{eq:boundary}, we have $\mathsf D_{\ell k}+\mathsf D_{k\ell}^*=u_\ell(0)^*u_k(0)$, and setting $\ell=k$ gives
\[
 |u_\ell(0)|^2=2\re\widehat\mu_\ell=\nu_\ell^2>0.
\]
Multiplying each $u_\ell$ by a unimodular constant, we may arrange that $u_\ell(0)=\nu_\ell$ while preserving orthonormality and the upper-triangular form of $\mathsf D$. Then $\mathsf D_{k\ell}=0$ for $\ell<k$, and \eqref{eq:boundary} gives the strictly upper-triangular entries
\[
 \mathsf D_{\ell k}=u_\ell(0)^*u_k(0)=\nu_\ell\nu_k\qquad(\ell<k).
\]
With $\mathsf W_b=\diag(\nu_1,\ldots,\nu_{2M})$, the Hermitian matrix $\mathsf J=(\mathsf D-\mathsf D^\dagger)/\ii$ representing $2\im\mathcal D$ takes the form
\[
 \mathsf J=\mathsf W_b\bigl[\diag(\widehat a_\ell/\nu_\ell^2)+\mathsf S_{2M}\bigr]\mathsf W_b,
\]
while $\mathcal E$ is represented by the row vector $\mathbf1^{\mathsf T}\mathsf W_b$.

We show that for a fixed $\sigma\in\{1,-1\}$, the quadratic form $\sigma\langle g,(\im\mathcal D)g\rangle$ is positive definite on $\ker\mathcal E$ if and only if $\sigma a_j>0$ ($1\le j\le M$) and \eqref{eq:standardarctan} holds. For $g=\sum_{\ell=1}^{2M}z_\ell u_\ell$ with $z\in\C^{2M}$, the condition $g\in\ker\mathcal E$ reads $\mathbf1^{\mathsf T}(\mathsf W_bz)=0$, i.e., $\mathsf W_bz\in\mathbf1^\perp$. Under the substitution $\xi=\mathsf W_bz$ (followed by complex conjugation when $\sigma=-1$), the positive definiteness of $\sigma z^\dagger\mathsf Jz$ on $\ker\mathcal E$ is equivalent to that of the compression of $\diag(c_1,\ldots,c_{2M})+\mathsf S_{2M}$ to $\mathbf1^\perp$, where $c_{2j-1}=c_{2j}=\sigma a_j/b_j$.

For an eigenvector $\xi\in\mathbf1^\perp$ of this compression with eigenvalue $\omega\in\R$, set $p_0=0$ and $p_\ell=\sum_{k=1}^\ell\xi_k$. Then $p_{2M}=0$ and
\[
 (c_\ell-\omega+\ii)p_\ell-(c_\ell-\omega-\ii)p_{\ell-1}=\eta,
 \qquad 1\le\ell\le2M,
\]
for some $\eta\ne0$. Subtracting $\eta/(2\ii)$ from each $p_\ell$ yields a homogeneous recurrence with equal nonzero endpoint values. Conversely, any such solution gives an eigenvector $\xi\in\mathbf1^\perp$. Thus $\omega$ is an eigenvalue precisely when
\[
 \prod_{\ell=1}^{2M}\frac{c_\ell-\omega-\ii}{c_\ell-\omega+\ii}=1.
\]
With the branch $\arg(c_\ell-\omega+\ii)\in(0,\pi)$, the sum $\sum_{\ell=1}^{2M}\arg(c_\ell-\omega+\ii)$ increases strictly from $0$ to $2M\pi$ as $\omega$ runs from $-\infty$ to $+\infty$. The smallest eigenvalue $\omega$ is the value at which this sum equals $\pi$, so the compression is positive definite precisely when the sum at $\omega=0$ is strictly less than $\pi$. Since the $c_\ell$ occur in pairs, this requires $c_\ell>0$ for all $\ell$; then $\arg(c_\ell+\ii)=\arctan(1/c_\ell)$, which gives \eqref{eq:standardarctan}.

Under the hypotheses of Theorem~\ref{th:regularity}(i), taking $\sigma=\sgn(a_1)$ therefore gives $\sigma\langle g,(\im\mathcal D)g\rangle>0$ for all nonzero $g\in\ker\mathcal E$. Applying Lemma~\ref{lem:endpoint} to $\phi_1,\ldots,\phi_M$ and using \eqref{eq:bridge} gives
\[
 |f_{{\rm std},M}|\ge
 \frac{c_{\mathcal D}^M\det\mathsf B}{|\mathcal C_M|}>0 ,
\]
so $f_{{\rm std},M}$ has no zeros in $\R^3$.

It remains to identify the leading homogeneous part of $f_{{\rm std},M}$ in $(x,y)$. In the expansion of \eqref{eq:standard}, every non-identity permutation contains at least two constant off-diagonal entries and has degree at most $2M-2$ in $(x,y)$. The homogeneous part of highest degree $2M$ is therefore the product of the linear parts in $(x,y)$ of the diagonal entries,
\[
 f_{\rm lead}(x,y)=\prod_{j=1}^M|x+\lambda_jy|^2.
\]
Since $b_j>0$, $f_{\rm lead}$ is independent of $t$ and strictly positive on the unit circle $x^2+y^2=1$. Combining Theorem~\ref{th:construction}(i) with the results of Sect.~\ref{sec:decay} completes the proof. \qed
\end{proof}

\begin{remark}\label{rem:equality}
When $M\ge2$, $a_1,\ldots,a_M$ are all positive or all negative, and equality holds in \eqref{eq:standardarctan}, the conclusions of Theorem~\ref{th:regularity}(i) remain valid.
\end{remark}

\begin{proof}[of Theorem~\ref{th:regularity}\textup{(ii)}]
By the expansion preceding \eqref{eq:highergram}, the functions $\phi_1,\ldots,\phi_N$ belong to the $(d+1)$-dimensional $\mathcal D$-invariant space $\mathcal V_d=e^{\mu s}\mathbb P_d(s)$, where $\phi_i(s)e^{-\mu s}$ has degree $n_i$ and leading coefficient $1/(n_i!)$. Since $n_1<\cdots<n_N$, these functions have distinct degrees and are therefore linearly independent for all $(x,y,t)\in\R^3$ and any fixed phase parameters.

In $\mathcal V_d$, consider the Laguerre basis functions
\[
 \ell_k^{\rm Lag}(s)=\sqrt b\,e^{\mu s}L_k(-bs),\qquad 0\le k\le d,
\]
where $L_k$ is the Laguerre polynomial of degree $k$ \cite[Chap.~V]{Szego1975}. Since $2\re\mu=b$, substituting $\chi=-bs$ and using the orthonormality of the Laguerre polynomials yields
\[
 \langle\ell_j^{\rm Lag},\ell_k^{\rm Lag}\rangle
 =\int_0^\infty e^{-\chi}L_j(\chi)L_k(\chi)\dd\chi=\delta_{jk},
\]
so $\{\ell_k^{\rm Lag}\}_{k=0}^d$ is an orthonormal basis of $\mathcal V_d$. Using the identity $L_k'(\chi)=-\sum_{j<k}L_j(\chi)$, we have
\[
 \mathcal D\ell_k^{\rm Lag}
 =\mu\ell_k^{\rm Lag}-b\sqrt b\,e^{\mu s}L_k'(-bs)
 =\mu\ell_k^{\rm Lag}+b\sum_{j<k}\ell_j^{\rm Lag}.
\]
Thus $\mathcal D$ is represented in this basis by $\mathsf D_{\rm Lag}=\mu\mathsf I+b\mathsf T$, where $\mathsf T_{jk}=1$ for $j<k$ and $0$ otherwise, and $2\im\mathcal D$ is represented by
\[
 \frac{\mathsf D_{\rm Lag}-\mathsf D_{\rm Lag}^\dagger}{\ii}=a\mathsf I+b\mathsf S_{d+1},\qquad
 \mathsf S_{d+1}=\frac{\mathsf T-\mathsf T^{\mathsf T}}{\ii}.
\]

Since $L_k(0)=1$, the evaluation functional $\mathcal E$ is represented in this basis by the row vector $\sqrt b\,\mathbf1^{\mathsf T}$, so $\ker\mathcal E$ corresponds to $\mathbf1^\perp$. Under condition~\eqref{eq:cotbound}, we have $a\ne0$; with $\sigma=\sgn(a)$, Lemma~\ref{lem:signspectrum} shows that the compression of $\sigma(a\mathsf I+b\mathsf S_{d+1})/2$ to $\mathbf1^\perp$ has smallest eigenvalue
\[
 \frac12\left(|a|-b\cot\frac{\pi}{d+1}\right)>0,
\]
so $\sigma\langle g,(\im\mathcal D)g\rangle>0$ for all nonzero $g\in\ker\mathcal E$. Applying Lemma~\ref{lem:endpoint} with $\mathsf M=\mathsf P\mathsf H\mathsf P^\dagger$ from \eqref{eq:highertau} then gives $|f_\Lambda|\ge c_{\mathcal D}^N\det\mathsf B>0$, showing that $f_\Lambda$ has no zeros in $\R^3$.

To find the highest-degree homogeneous part of $f_\Lambda$ in $(x,y)$ at fixed $t$, note from \eqref{eq:physicalschur} that the linear homogeneous part of $s_1$ in $(x,y)$ is $w_0=x+\lambda^*y$, whereas $s_2$ has degree $1$ in $(x,y)$ and $s_k$ is independent of $(x,y)$ for $k\ge3$. Assigning weight $k$ to $s_k$ in \eqref{eq:schurseries} makes $S_m(\boldsymbol s)$ weighted homogeneous of weight $m$; since the degree of $s_k$ in $(x,y)$ is strictly less than $k$ for $k\ge2$, the highest-degree homogeneous part of $S_m(\boldsymbol s)$ in $(x,y)$ comes solely from $s_1^m$ and equals $w_0^m/m!$.

Set
\[
 D_\Lambda=\sum_{i=1}^N n_i-\frac{N(N-1)}2.
\]
For any strictly increasing column index tuple $I=(k_1<\cdots<k_N)\subset\{0,\ldots,d\}$, let $\mathsf P_{:,I}$ denote the submatrix of $\mathsf P$ with columns indexed by $I$. Each nonzero term in the minor $\det(\mathsf P_{:,I})$ has weight $\sum_{i=1}^N n_i-\sum_{j=1}^N k_j$, which bounds its degree in $(x,y)$. Since $k_j\ge j-1$, this weight attains its maximum $D_\Lambda$ if and only if $I=I_0=(0,1,\ldots,N-1)$, for which the leading homogeneous part of $\det(\mathsf P_{:,I_0})$ in $(x,y)$ is $K_\Lambda w_0^{D_\Lambda}$ (with $1/(n_i-k)!=0$ for $n_i<k$). Factoring $1/n_i!$ from the $i$th row reduces the $(i,k)$ entry to the monic degree-$k$ polynomial $n_i(n_i-1)\cdots(n_i-k+1)$ in $n_i$. By the Vandermonde determinant formula,
\[
 K_\Lambda=\det\left[\frac1{(n_i-k)!}\right]_{1\le i\le N,\,0\le k<N}=\frac{\prod_{1\le i<j\le N}(n_j-n_i)}{\prod_{i=1}^N n_i!}>0 .
\]

Applying the Cauchy--Binet formula twice to \eqref{eq:highertau} gives
\[
 f_\Lambda=
 \sum_{\substack{I,J\subset\{0,\ldots,d\}\\ |I|=|J|=N}}
 \det(\mathsf P_{:,I})\,\det(\mathsf H_{I,J})\,
 \bigl[\det(\mathsf P_{:,J})\bigr]^* .
\]
Since the constant matrix $\mathsf H$ is independent of $(x,y,t)$, only the pair $(I,J)=(I_0,I_0)$ can contribute terms of degree $2D_\Lambda$ in $(x,y)$. Comparing the coefficients of $\epsilon^k\delta^\ell$ in $(q+\delta)/(b+\epsilon+\delta)=\int_{-\infty}^0(q+\delta)e^{(b+\epsilon+\delta)s}\dd s$, we write $\mathsf H_{k\ell}$ as the weighted inner product
\[
 \mathsf H_{k\ell}
 =\left\langle \frac{s^k}{k!},(q+\partial_s)\frac{s^\ell}{\ell!}\right\rangle_b,
 \qquad
 \langle \varphi_1,\varphi_2\rangle_b=\int_{-\infty}^{0}
      e^{bs}\varphi_1^*(s)\varphi_2(s)\dd s.
\]
The orthogonal polynomials
\[
 p_k(s)=b^{-k}L_k(-bs),\qquad \langle p_k,p_k\rangle_b=b^{-2k-1}\qquad(0\le k\le N-1),
\]
have the same leading coefficient $1/k!$ as $s^k/k!$, so the change-of-basis matrix between the two bases is unit triangular, leaving $\det(\mathsf H_{I_0,I_0})$ unchanged. Since $\partial_s p_\ell\in\operatorname{span}\{p_0,\ldots,p_{\ell-1}\}$ is orthogonal to $p_k$ for $k\ge\ell$, the weighted inner product matrix $\bigl[\langle p_k,(q+\partial_s)p_\ell\rangle_b\bigr]_{k,\ell=0}^{N-1}$ is upper triangular with diagonal entries $q\langle p_k,p_k\rangle_b=q\,b^{-2k-1}$ ($0\le k\le N-1$). Using $\sum_{k=0}^{N-1}(2k+1)=N^2$, we find
\[
 \det(\mathsf H_{I_0,I_0})=\prod_{k=0}^{N-1}q\,b^{-2k-1}
 =\frac{q^N}{b^{N^2}},
\]
which yields the spatial asymptotic expansion
\begin{equation}\label{eq:spatialleading}
 f_\Lambda=\frac{q^N}{b^{N^2}}K_\Lambda^2|w_0|^{2D_\Lambda}
                 +O(R^{2D_\Lambda-1}).
\end{equation}
Since $b>0$, we have $q\ne0$, and $|w_0|=|x+\lambda^*y|$ is strictly positive on the unit circle; hence the leading homogeneous term in \eqref{eq:spatialleading} is independent of $t$, has constant argument, and is bounded away from zero on the unit circle. Combining Theorem~\ref{th:construction}(ii) with the results of Sect.~\ref{sec:decay} completes the proof. \qed
\end{proof}
\end{theopargself}

\begin{remark}\label{rem:sharpness}
Condition~\eqref{eq:cotbound} in Theorem~\ref{th:regularity}(ii) is sharp in the sense that for any $d\ge1$ and $b>0$, the higher-order $\tau$-functions $f_\Lambda$ have no zeros in $\R^3$ for all index vectors $\Lambda$ with $n_N=d$ and all phase parameters $\zeta_k$ if and only if $|a|>b\cot(\pi/(d+1))$. The case $d=1$ was verified in Sect.~\ref{sec:standard}. For $d\ge2$ and $|a|\le b\cot(\pi/(d+1))$, take the single-component index vector $\Lambda=(d)$. For $r^{d+1}=1$ and $r\ne1$, let $g_r=\sum_{k=0}^d r^k\ell_k^{\rm Lag}$. These are the modes corresponding to the compression in Lemma~\ref{lem:signspectrum}. Each has a polynomial part of degree $d$, and
\[
 g_r(0)=0,\qquad g_r'(0)=-\frac{(d+1)b^{3/2}}{1-r}\ne0.
\]

At equality, take the extreme mode on which $\im\mathcal D$ vanishes; below the bound, combine the two orthogonal extreme modes (on which $\im\mathcal D$ has opposite signs) with a suitable amplitude ratio and relative phase so that $\langle g,(\im\mathcal D)g\rangle=0$ while the leading coefficient of $e^{-\mu s}g(s)$ and $g'(0)$ are both nonzero. By \eqref{eq:boundary}, $\langle g,\mathcal Dg\rangle=0$. Scaling $g$ so that $e^{-\mu s}g(s)$ has leading term $s^d/d!$, writing $f=f_{(d)}$ and $\phi=\phi_1$, and choosing $\zeta_1,\ldots,\zeta_d\in\C$ such that $\phi|_{x=y=t=0}=g$ gives $f(0,0,0)=\langle\mathcal Dg,g\rangle=0$. Since $\phi_x=(\mathcal D-\mu)\phi$ and $\phi_y=-\ii(\mathcal D^2-\mu^2)\phi$, integration by parts yields, at $(0,0,0)$,
\[
 f_x=0,\qquad f_{xx}=|g'(0)|^2>0,\qquad f_y=-\ii|g'(0)|^2\ne0,
\]
so along the parabola $t=0$, $y=x^2/2$ we have $v(x,x^2/2,0)=4/x+O(1)$ as $x\to0$.
\end{remark}

\section{Evolution of lump solutions}\label{sec:visualization}

\begin{figure}[!htbp]
\centering
\includegraphics[width=.49\linewidth]{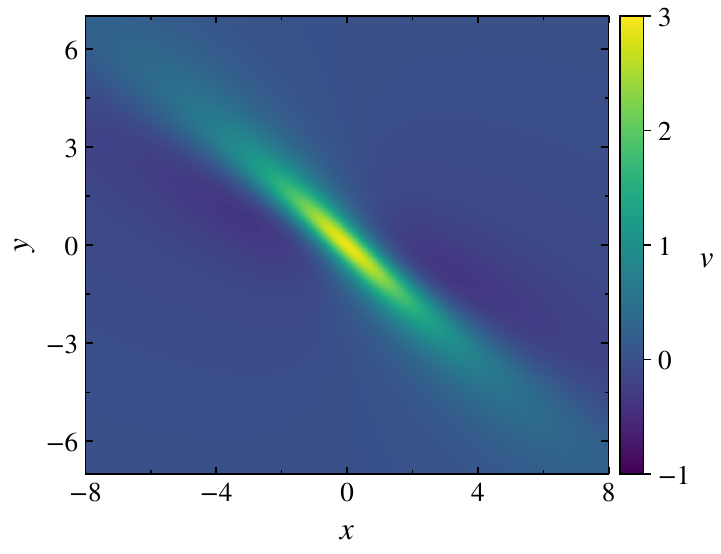}
\caption{Density plot of the zero-phase fundamental lump solution at $t=0$ ($a=1.17$, $b=0.65$).}\label{fig:localization}
\end{figure}

We now illustrate Theorem~\ref{th:regularity} with the fundamental lump and two higher-order lump solutions. Figure~\ref{fig:localization} plots the zero-phase ($\gamma_1=0$) fundamental lump solution \eqref{eq:fundamental} at $t=0$ for $a=1.17$ and $b=0.65$. The solution has a positive peak at the origin and two symmetric negative troughs.

\begin{figure}[!htbp]
\centering
\panel{.325}{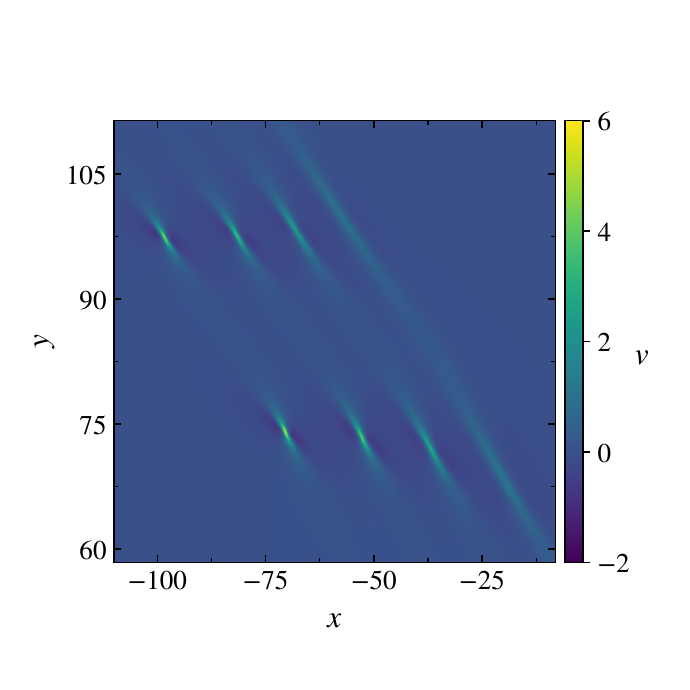}{(a) $t=-12$}\hfill
\panel{.325}{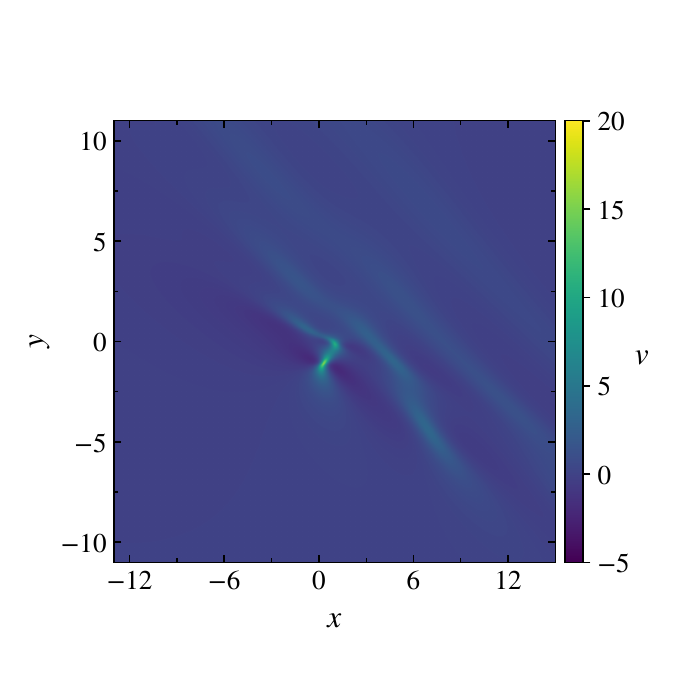}{(b) $t=0$}\hfill
\panel{.325}{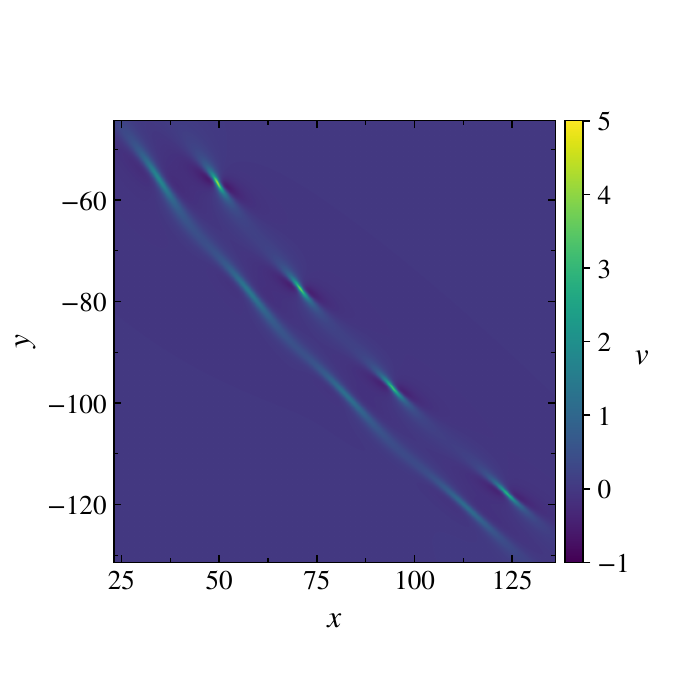}{(c) $t=12$}
\caption{Evolution of the zero-phase higher-order lump solution $v$ for $\Lambda=(2,3,4,5)$ and $(a,b)=(1.20,0.65)$.}\label{fig:finiteA}
\end{figure}

Figure~\ref{fig:finiteA} shows the evolution of the zero-phase ($\zeta_1=\ldots=\zeta_d=0$) higher-order lump solution for $\Lambda=(2,3,4,5)$ and $(a,b)=(1.20,0.65)$; here $d=5$, $D_\Lambda=8$, and $b\cot(\pi/6)\approx 1.1258<1.20=|a|$, so~\eqref{eq:cotbound} holds. At $t=-12$ (Fig.~\ref{fig:finiteA}(a)), eight peaks appear in two rows; they merge near the origin at $t=0$ with increased peak amplitudes (Fig.~\ref{fig:finiteA}(b)), and then separate into four pairs at $t=12$, exhibiting anomalous scattering (Fig.~\ref{fig:finiteA}(c)).

\begin{figure}[!htbp]
\centering
\panel{.325}{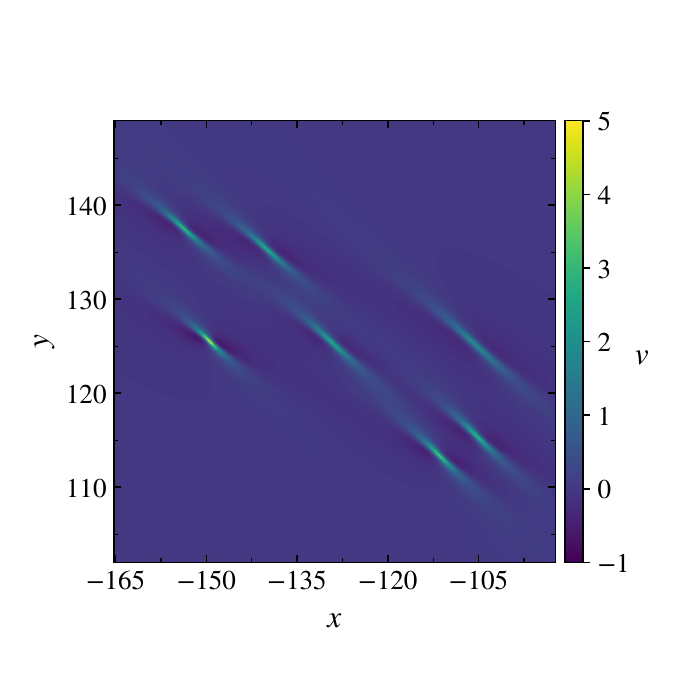}{(a) $t=-12$}\hfill
\panel{.325}{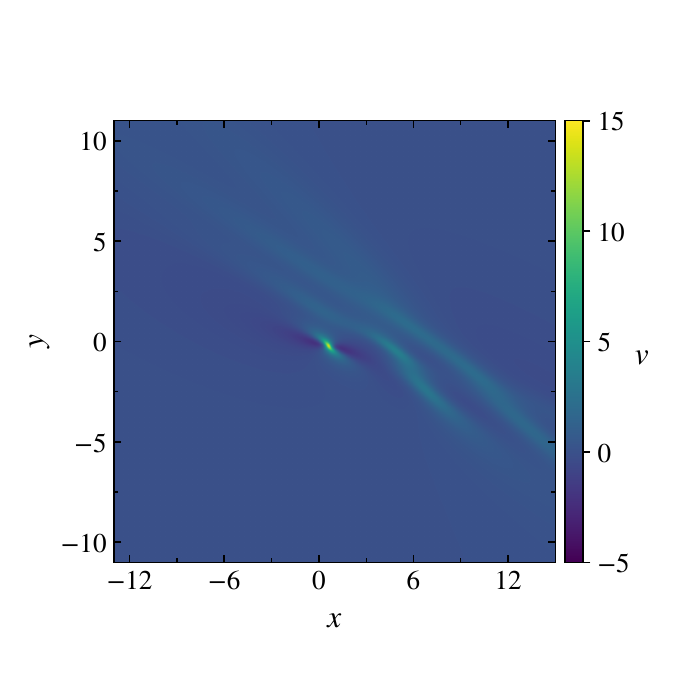}{(b) $t=0$}\hfill
\panel{.325}{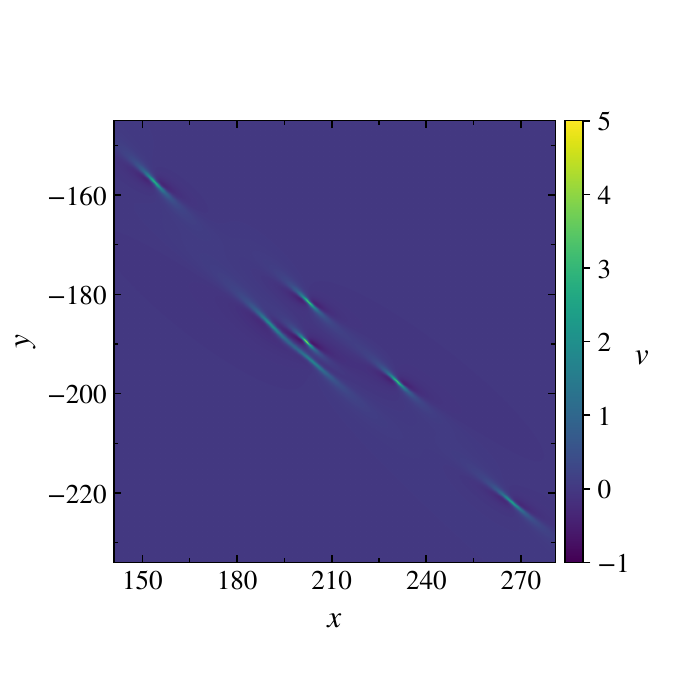}{(c) $t=18$}
\caption{Evolution of the zero-phase higher-order lump solution $v$ for $\Lambda=(1,2,4,6)$ and $(a,b)=(1.75,0.80)$.}\label{fig:finiteB}
\end{figure}

As a second example, Fig.~\ref{fig:finiteB} displays the zero-phase solution with $\Lambda=(1,2,4,6)$ and $(a,b)=(1.75,0.80)$, for which $d=6$, $D_\Lambda=7$, and $b\cot(\pi/7)\approx 1.6612<1.75=|a|$, again satisfying~\eqref{eq:cotbound}. Seven peaks are visible at $t=-12$ (Fig.~\ref{fig:finiteB}(a)); following a central interaction near $t=0$ (Fig.~\ref{fig:finiteB}(b)), five peaks cluster near the center at $t=18$ while two outer peaks lie on opposite sides (Fig.~\ref{fig:finiteB}(c)).

\FloatBarrier

\medskip
\noindent \textbf{Acknowledgements} This work is supported by the Kaiyuan International Mathematical Sciences Institute, the Fundamental Research Funds for the Central Universities (JKF2025077492962), and the Academic Excellence Foundation of BUAA for Ph.D. Students.

\section*{Declaration of generative AI use}
During the preparation of this manuscript, the authors used GPT-6.1 Sol, GPT-6 Astra, and Gemini 4 Argon to assist with mathematical derivations. They reviewed, verified, and edited all AI-generated content and take full responsibility for its correctness and for the final manuscript.

\section*{Declarations}
\noindent \textbf{Conflict of interest} The authors declare that they have no conflict of interest.

\noindent\textbf{Data availability} The computational materials supporting this work are not publicly available. Requests for further information should be addressed to the authors.

\end{sloppypar}
\end{document}